# Low-angle misorientation dependence of the optical properties of InGaAs/InAlAs quantum wells.

Robert J. Young[*,1], Lorenzo O. Mereni[1], Nikolay Petkov[1], Gabrielle R. Knight[1], Valeria Dimastrodonato[1], Paul K. Hurley[1,] Greg Hughes[2] and Emanuele Pelucchi[1]

[1.]*Tyndall National Institute-University College Cork, Lee Maltings, Cork, Ireland*

[2.]*School of Physical Sciences, Dublin City University, Glasnevin, Dublin 9, Ireland*

**Abstract**

We investigate the dependence of the low-temperature photoluminescence linewidths from InP-lattice-matched InGaAs/InAlAs quantum wells on the low-angle misorientation from the (100) surface of the host InP substrate. Quantum wells were grown on InP substrates misorientated by 0, 0.2, 0.4 and 0.6°; 0.4° was found to consistently result in the narrowest peaks, with the optimal spectral purity of ~4.25 meV found from a 15nm quantum well. The width of the emission from the 15nm quantum well was used to optimize the growth parameters. Thick layers of Si-doped InGaAs were then grown and found to have bulk, low temperature (77K), electron mobilities up to $\mu \sim 3.5 \times 10^4$ cm$^2$/Vs with an electron concentration of ~1 x $10^{16}$.

---

[*] Corresponding author, email: jocg@r-j-y.com, phone: +44 7772 312557, current address : Department of Physics, Lancaster University, Lancaster, LA1 4YW, UK.

## 1. Introduction

Epitaxially grown (Al)InGaAs layers lattice matched to InP are of major importance for a wide variety of applications, such as in the active layers for telecom wavelength optoelectronic devices[1], quantum cascade lasers[2] and they are currently being investigated as potential high-mobility channel layers in future CMOS technology generations.[3,4] Recent metal organic vapor phase epitaxy (MOVPE) growth studies of (Al)GaAs on GaAs have shown that small misorientations in the range 0 to 0.6° on the GaAs substrate can have a dramatic impact in the optical and transport properties of the grown layers, allowing record material properties to be achieved for MOVPE, with results comparable with those obtainable by the best MBE systems.[5,6] These improvements have been shown to correlate with the substrate surface ordering, where the best optical properties were obtained from growth on substrates intentionally miscut at 0.2° off (100), which corresponds to the onset of the appearance of step bunching on the GaAs surface.[5,7,8] A similar study for growth of InGaAs on low angle misorientated InP substrates has not been reported, and is the subject of this paper.

In this work we investigate the effect of small substrate misorientation on the optical and electronic properties of InGaAs structures grown on InP. A layer structure containing three InP lattice matched InGaAs/InAlAs quantum wells was grown by MOVPE on a series of InP substrates misoriented from (100) by 0, 0.2, 0.4 and 0.6° towards the (111)A planes. Initially the photoluminescence emission width of the widest well was used as a quality indicator to optimize growth parameters, namely temperature, V/III ratio, alloy composition and growth rate. Following optimization, bulk layers of Si-doped InGaAs were grown on InP to assess the electron mobility.

## 2. Experimental

All samples were grown in an Aixtron 200/4 horizontal MOVPE reactor at 20mbar using trimethylgallium, trimethylindium and trimethylaluminium group-III sources, purified arsine and phosphine group-V sources, silane as an n-type dopant and purified nitrogen for the carrier gas. Constant checks of the reactor status/quality were performed by routine growth of thick GaAs QWs with AlGaAs barriers as in Ref 5. More details can be found in Ref. 6. We systematically obtain QW's with < 1meV linewidths, confirming state of the art MOVPE growth quality.[9]

For optical studies the following sample design was grown: 300nm InP on (100) misorientated InP substrates followed by three $In_{0.53}Ga_{0.47}As$ quantum wells, approximately lattice matched to InP with nominal widths of 2, 5 and 15nm, each surrounded by 300nm $In_{0.52}Al_{0.48}As$ barriers (compositions quoted are nominal). This was capped with 10nm of $In_{0.53}Ga_{0.47}As$ to minimize surface oxide formation. This sample structure is illustrated in fig. 1(a).

For electrical characterization a single 250nm layer of InP lattice matched n-doped $In_{0.53}Ga_{0.47}As$ was grown on InP substrates at three different doping concentrations (~1 x $10^{16}$, 1 x $10^{17}$ and 1 x $10^{18}$/cm$^3$). Hall measurements in the Van der Pauw geometry were made on these samples to assess the doping levels and mobilities both at room temperature and at 77K.

All structures were grown simultaneously on four semi-insulating InP substrates nominally misoriented from (100) by 0, 0.2, 0.4 and 0.6° ± 0.02°, allowing results from the different misorientations to be compared directly.

An atomic force microscope (AFM) in non-contact mode was used both to assess the surface morphology of samples post growth and also to investigate the grown structure in cross section (xAFM). Height contrast in xAFM is provided by the spontaneous growth of an oxide (~0.5nm thick) on aluminium rich layers following cleaving in air. This technique allowed the thickness of grown layers to be measured with a precision a few nm's. A sample xAFM image of the triple quantum well (TQW) structure for optical analysis is shown in fig. 1(b).

The composition of the two ternary alloys used was measured and adjusted to be close to the InP lattice match composition using a computer controlled Philips X'Pert X-ray diffractometer with a Bragg scan of the (400) reflection. Photoluminescence measurements were made at 10K in a closed-cycle Helium cryostat using a CW 650nm diode laser for excitation and a high resolution spectrometer fitted with a liquid nitrogen cooled InGaAs array for wavelength resolved spectroscopy.

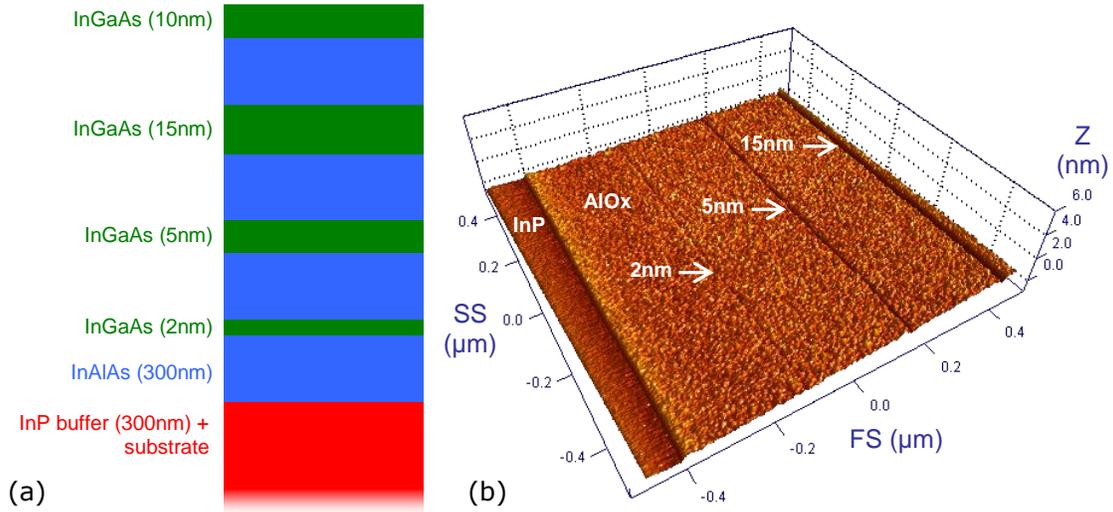

Fig. 1: (a) An illustration of the undoped sample design which consists of three, close to InP lattice matched, InGaAs quantum wells of indicated thickness embedded in 300nm InAlAs barriers. (b) Cross-sectional image of the grown quantum wells taken with an Atomic Force Microscope; the Fast Scan FS (Slow Scan SS) axis is parallel (perpendicular) to the growth direction.

## 3. Results and discussion

An AFM was used to study the surface morphology of our MOVPE grown samples. Fig. 2 shows AFM amplitude images, chosen to highlight the terrace step edges, of the surface of the TQW structure on each of the four misorientations of InP substrate. Where possible, the number of step edges in a given dimension was used to evaluate the true misorientation of each of the substrates from (100) and these were found to be within the manufacturer's tolerance of ± 0.02° except for the 0° wafer where the local misorientation was found to vary slightly outside of this tolerance; the local misorientation in fig. 2(a) is 0.026°. Between substrate misorientations of 0° and 0.4° there is a clear transition in the surface morphology from single terrace step flow to a complex step bunching, analogous to what has been reported for the GaAs system, despite the differences of lattice constants, composition and surface adatom mobilites.[5]

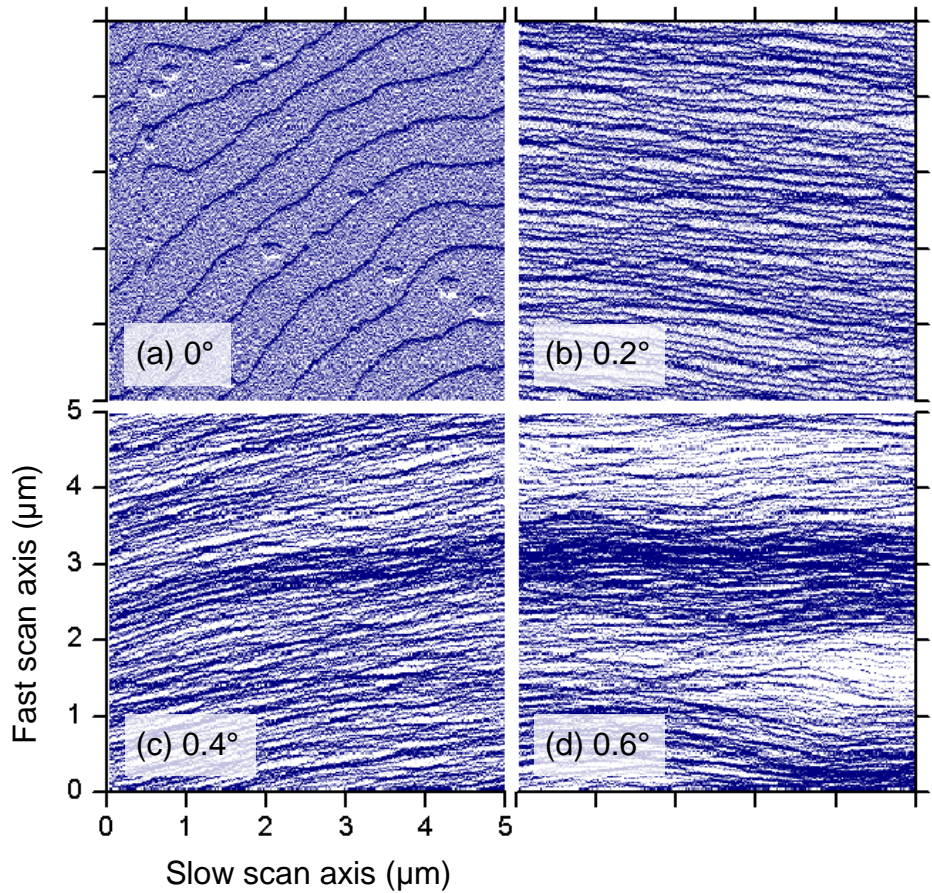

Fig. 2: Atomic Force Microscope amplitude images of the surface of the samples following growth of the quantum well structure illustrated in Fig. 1. Panels (a)-(d) correspond to growth on InP wafers misorientated by 0, 0.2, 0.4 and 0.6° from (100) respectively. The color scale is arbitrary and was chosen to highlight terrace edges.

To investigate the impact of the substrate misorientation on the optical properties of the TQW, we systematically studied the PL spectra as a function of the substrate miscut. A typical room temperature photoluminescence spectrum from a TQW sample is shown in fig. 3. The widths of the emission peaks from each of the three wells are all considerably broader than their quantum limits. A number of mechanisms contribute to the broadening of low temperature excitonic photoluminescence from quantum wells. Interface roughness, alloy disorder (local modulations in the composition of the alloys) in both the barrier and well materials and the unwanted incorporation of impurities in the lattice, all act to perturb the excitons' confinement energies and hence inhomogeneously broaden their emission (other effects can also play significant roles in broadening emission, including; spectral meandering induced by local electric fields associated with free charges, phonon coupling and energetic level repulsion etc.).[10-14]

The emission linewidth was chosen as a figure of merit, in order to optimize the growth conditions to bring the observed photoluminescence from the quantum wells as close to their quantum limit as possible. The small peak to the low energy side of the 15nm well's main peak is due to recombination from a second exciton species (an exciton with a different electronic configuration) in the well, whose emission energy is perturbed by the Coulomb interaction.[15] Such peaks are typically unresolved and add to the apparent line width of the well. Where resolvable, peaks were fitted with a double Gaussian lineshape to extract the linewidth of a single exciton species.

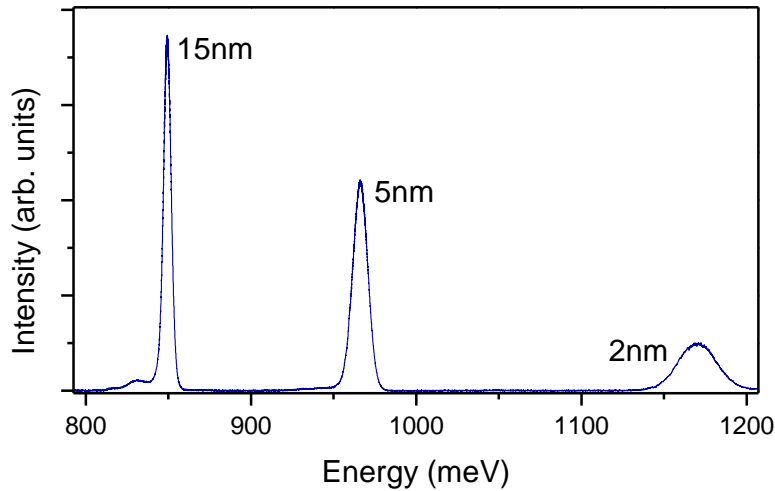

Fig. 3: A typical photoluminescence spectrum taken from the quantum well structure illustration in Fig. 1.

In a series of six growths the conditions under which the TQW structure was grown were optimized to minimize the photoluminescence linewidths of the wells. Fig. 4 (a) shows the width of the emission from the 15nm well as a function of the growth run, as the growth conditions were tuned the linewidth of the well reduced by a factor of ~2. The optimal growth conditions we found were a growth temperature of 710°C (thermocouple, ~650°C actual) with a constant V/III ratio of 400, through both the well and barrier layers.

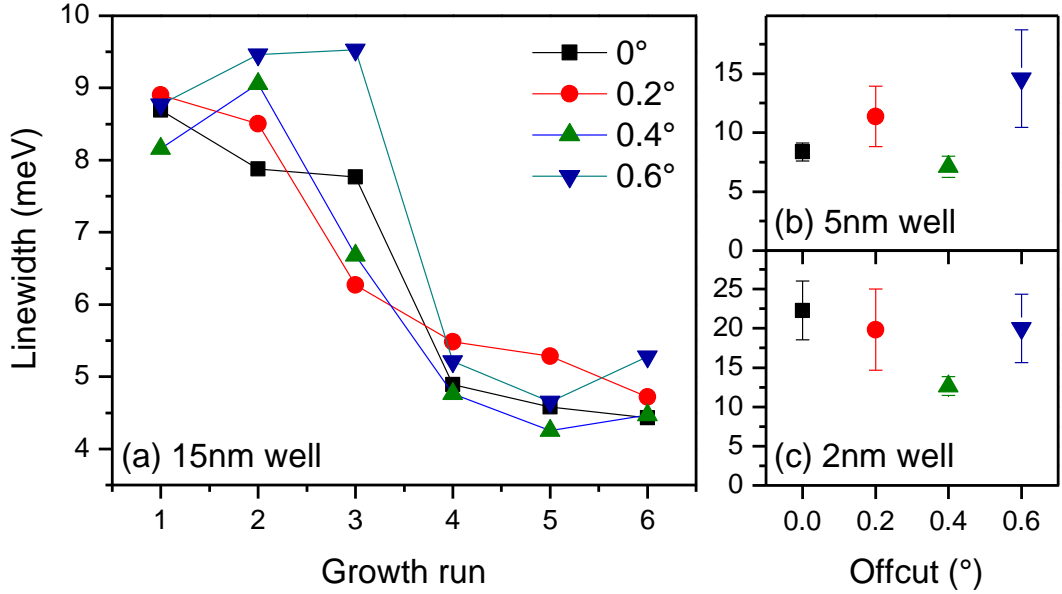

Fig. 4: (a) The linewidth of the photoluminescence emission from the 15nm quantum well for six different growth runs on four different InP wafer misorientations from (100) as labeled. (b) and (c) The average linewidth for all six growth runs plotted as a function of the substrate misorientation for the 5nm and 2nm wells respectively.

Following optimization of the growth conditions it is clear from fig. 4 (a) that the emission from the 15nm well was significantly narrower (with a minimum linewidth of 4.25 meV, slightly lower than the best reports to date for this system, irrespective of the growth method)[16] from the 0° and 0.4° miscut substrates than from the 0.2° and 0.6° substrates. A transmission electron microscopy micrograph illustrating the good quality interfaces of the 2nm quantum well is shown in fig. 5.

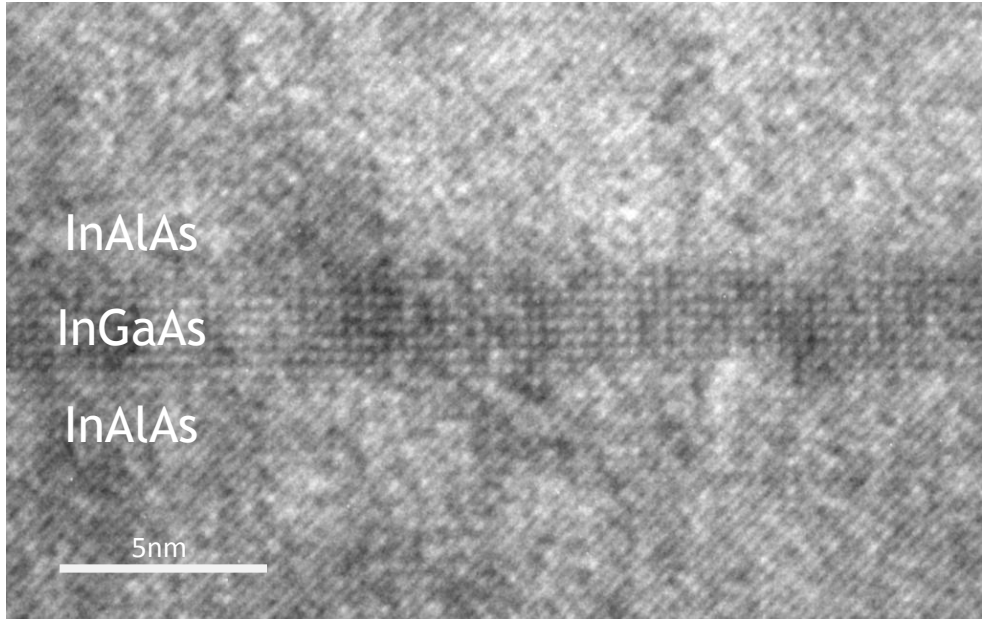

Fig. 5: A transmission electron microscopy micrograph of a 2nm InGaAs/InAlAs quantum well grown on an InP substrate misorientated from (100) by 0.4°.

The average width of the well emission for all six growth for the 5nm and 2nm quantum wells are shown in fig. 4 (b) and (c) respectively. Results from the 5nm InGaAs QW closely resemble those from the 15nm well, but while the 0.4° offcut gave the narrowest emission from the 2nm well consistently, the 0° offcut surprisingly showed the broadest emission. Error bars in fig. 4 (b) and (c) span two standard deviations from the mean, the magnitude of the bars appears to be correlated to the width of the wells' emission. This may be due to multiple exciton-species' peaks being present in the emission which are resolved (or not present) for wells with narrow emission widths, and unresolved and hence included, in the linewidth for wells with broad emission widths. For the broadest emission lines measured, for the first three growth runs in Fig.4 (a), there is considerable 'switching' in the misorientation that shows the narrowest emission. Small, unresolved contributions from two or more differing exciton species of varying relative intensity are also thought to be the cause of this switching. For more optimized growths, where the linewidth of emission has narrowed significantly, emission from differing exciton species can be resolved and measuring the true inhomogeneous linewidth from the neutral exciton becomes reliable and, in-turn, the results become consistent.

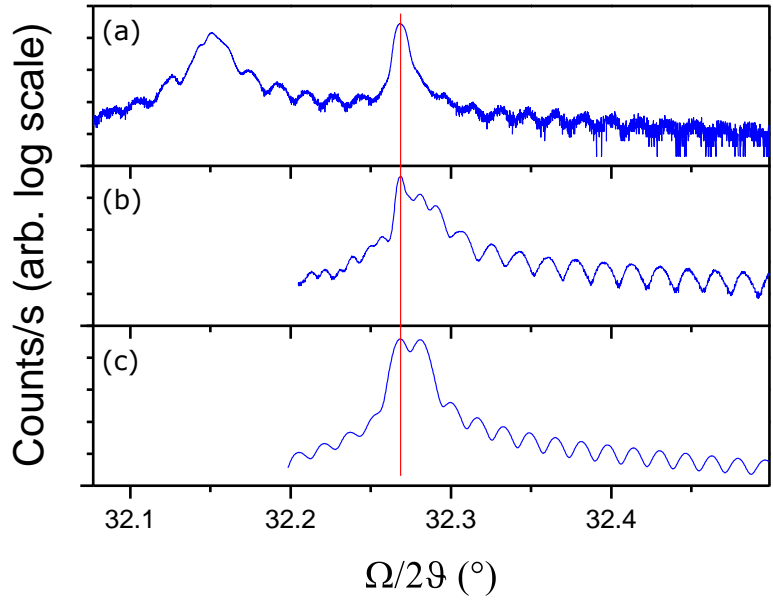

Fig. 6: X-ray diffraction spectra taken from (a) the initial, and (b) optimized, growth of the structure illustrated in Fig. 1. Panel (c) shows the expected ideal spectrum for the sample design.

A (400) $\Omega-2\theta$ x-ray diffraction spectrum measured from the optimized TQW structure is shown in Fig. 6(b) alongside a simulation of the structure in 6(c). The two spectra are quite similar, in contrast to a spectrum taken from the un-optimized TQW structure shown in fig. 6(a). Grown at a lower temperature (690°C, thermocouple) and V/III ratio (200) the un-optimized structure has a layer peak with a slight lattice mismatch and considerable width, and the poor visibility of the Pendellösung fringes, which is indicative of low quality hetero-interfaces, is much higher than for the optimized structure. The improvement in PL linewidth with the optimization of the growth parameters, seen, in fig. 4(a), is thus attributed to both an improvement in the purity of the ternary compound in the well (reducing alloy scattering) and also an improvement in the well/barrier interface abruptness (reducing interface scattering and the inhomogenous broadening associated with a 1D well of varying width).

Bulk Hall mobilities for the three samples grown for electrical characterization, after optimizing growth conditions for InGaAs, are shown in fig. 7 as a function of the measured carrier density. The room temperature mobility is limited by phonons and alloy scattering,[17] and is roughly equal for all samples. At low temperature (77K) phonon scattering becomes less important and impurity/alloy scattering dominates leading to an exponential increase in mobility as a function of decreasing carrier density, with results in line with what is reported in the literature for these systems.[18] Within experimental error no dependence between the substrate miscut and the resulting mobility was measured, indicating perhaps that the relationship highlighted with the optical study is a result of small changes in the background doping, or changes to the interfaces between the quantum wells and their barriers, or, for example, that the short range alloy disorder dependence on the substrate miscut affects optical properties of the QW's more than the bulk transport properties.

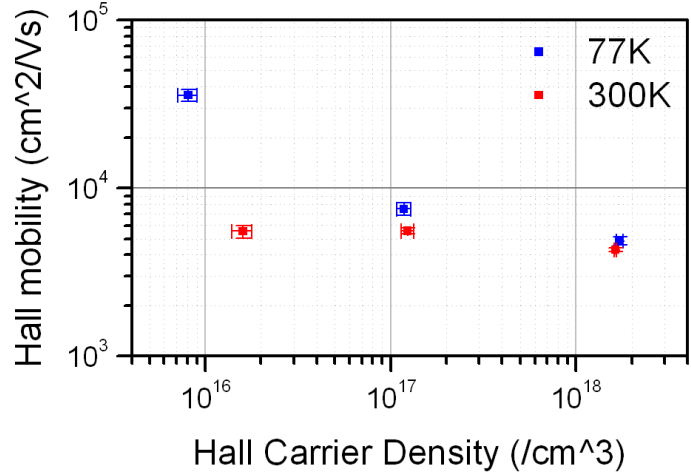

Fig. 7: The Hall mobility as a function of the carrier density of a bulk Si-doped InGaAs layer grown on InP using the optimized growth conditions found using the spectral purity of emission from a 15nm InGaAs quantum well as the figure of merit.

**4. Conclusions**

An InP lattice matched InGaAs/AlGaAs TQW structure was grown on substrates misorientated from (100) by small angles. The widths of the quantum well emission in low temperature, high resolution photoluminescence measurements were used as quality indicators to feedback small adjustments to the growth parameters. Iterating this approach, following six growth cycles the PL linewidth of the 15nm well was reduced by more than a factor of 2 to ~4.25 meV, the smallest value yet to be reported in this material system.[16] After optimization of the growth parameters the emission from the three wells grown on the substrate miscut by 0.4° were found to be consistently narrower than for other angles. Surfaces studies by AFM indicate that 0.4° degrees off surface develops a step-bunching profile.[5] Overall, this result, that the highest quality growth of quantum wells is achieved in an interval between the transition between step-flow and heavy step-bunching is consistent with the result from a previous growth study of GaAs quantum wells,[5] where the transition and minimum linewidths were found from samples grown on substrates misorientated by 0.2°. 0.4° appears to be the optimum small-angle misorientations for growth on InP substrates by MOVPE.

Electronic characterization of bulk n-doped InGaAs layers grown using the optimized conditions revealed mobilities up to $3.5 \times 10^4$ cm$^2$/Vs at 77K.

**Acknowledgements**

This research was enabled by the Irish Higher Education Authority Program for Research in Third Level Institutions (2007-2011) via the INSPIRE programme, and by Science Foundation Ireland under grants 05/IN.1/I25, 08/RFP/MTR1659, 05/IN.1/I25/EC07, 07/SRC/I1173 and the Ireland National Access Program at the Tyndall National Institute (No. 152). We are grateful to K. Thomas for his support with the MOVPE system.